\documentclass[11pt]{article}
\usepackage[normalem]{ulem}
\pdfoutput=1 

\usepackage{jheppub} 
\usepackage{url}
\usepackage{caption}
\usepackage{subcaption}
\usepackage{extarrows}

\usepackage[latin9]{inputenc}
\usepackage{float}
\usepackage{amsmath}
\usepackage{amssymb}
\usepackage{graphicx}
\usepackage{esint}
\usepackage{hyperref}
\usepackage{comment}
\usepackage{color}
\usepackage{microtype}
\usepackage{cleveref}
\usepackage{breakurl}
\usepackage{bbm}
\newcommand{\be}{\begin{equation}}
\newcommand{\ee}{\end{equation}}
\newcommand{\ben}{\begin{displaymath}}
\newcommand{\een}{\end{displaymath}}
\newcommand{\bea}{\begin{eqnarray}}
\newcommand{\eea}{\end{eqnarray}}
\def\K{K{\"a}hler }
   \newcommand{\rf}[1]{(\ref{#1})}
\newcommand{\vp}{\varphi}

\def\be{\begin{equation}}
\def\ee{\end{equation}}
\def\bea{\begin{eqnarray}}
\def\eea{\end{eqnarray}}
\def\ba{\begin{array}}
\def\ea{\end{array}}
\def\bit{\begin{itemize}}
\def\eit{\end{itemize}}

\def\a{\alpha}
\def\b{\beta}

\def\vp{\varphi}
\def\vt{\vartheta}

 \makeatletter

\allowdisplaybreaks

\makeatother

\makeatletter
\DeclareRobustCommand{\rcite}[1]{%
  \rcite@aux#1,\@nil{#1}%
}
\def\rcite@aux#1,#2\@nil#3{%
  \if\relax#2\relax
    Ref.~\cite{#3}%
  \else
    Refs.~\cite{#3}%
  \fi
}
\makeatother

\hypersetup{
    colorlinks = true,
    citecolor = {blue},
    linkcolor = {blue},
    urlcolor = {blue},
}

 \title{\rm {\bf \LARGE \boldmath  {$SL(2,\mathbb{Z})$   Cosmological   Attractors }}}

\author{Renata Kallosh and }
\author{Andrei Linde}

\affiliation{Stanford Institute for Theoretical Physics and Department of Physics,\\ Stanford University, Stanford, CA 94305, USA}
\emailAdd{kallosh@stanford.edu}
\emailAdd{alinde@stanford.edu}

\abstract{We study cosmological theory where the kinetic term and potential have $SL(2,\mathbb{Z})$ symmetry.  Potentials have a plateau at large values of the inflaton field,  where the axion forms a flat direction. Due to the underlying hyperbolic geometry and special features of $SL(2,\mathbb{Z})$ potentials, the theory exhibits an $\alpha$-attractor behavior: its cosmological predictions are stable with respect to significant modifications of the $SL(2,\mathbb{Z})$ invariant potentials. We present a supersymmetric version of this theory in the framework of $\overline {D3}$ induced geometric inflation.  The choice of $\alpha$ is determined by underlying string compactification. For example, in a CY compactification with  $T^2$, one has $3\alpha=1$, the lowest discrete Poincar\'e disk target for LiteBIRD.
}

\begin{document}

\maketitle


\parskip 2pt

\section{Introduction}

It was proposed in \cite{Casas:2024jbw} to approach inflationary cosmology from the quantum gravity/string theory perspective and view the modular invariant effective potential $V(\tau, \bar \tau)$  as the $SL(2,\mathbb{Z})$ invariant which already includes all relevant corrections, such as higher powers of the curvature corrections to the action.\footnote{A particular example of such an  $SL(2,\mathbb{Z})$ invariant theory presented in \cite{Casas:2024jbw} is based on an additional assumption that the potential depends on modular invariant species scale and its derivatives.  We will not make this assumption in our paper.}

We are motivated by the ``target space modular invariance'', a general concept introduced in  \cite{Ferrara:1989bc}. There was a strong interest in modular invariant effective potentials and modular inflation over the years; see a list of references [34-49] given in  \cite{Casas:2024jbw}.
The concept of automorphic inflation, including modular inflation, in particular, $j$-inflation,  was introduced and developed in \cite{Schimmrigk:2014ica,Schimmrigk:2016bde,Schimmrigk:2021tlv}. The $SL(2,\mathbb{Z})$ invariant potentials were studied there, and the almost holomorphic modularity of CMB observables was established. A relevant mathematical background can be found in \cite{Schimmrigk:2016bde}; see refs. [40-56]  there.
The novel feature of the proposal made in \cite{Casas:2024jbw}  was the requirement that the modular invariant effective potential $V(\tau, \bar \tau)$  has a plateau at large values of  ${\rm Im}(\tau)$, which results in a long period of inflation.

In this paper, we will construct a large class of $SL(2,\mathbb{Z})$ invariant theories with plateau potentials and study their cosmological properties.  We will show that these models have features closely related to a specific class of the cosmological $\alpha$-attractor models called T-models \cite{Kallosh:2013hoa,Kallosh:2013yoa,Galante:2014ifa,Kallosh:2015zsa}.

Our models depend on a single complex field $\tau = \tau_1+i \tau_2$,
\be
{ {\cal L} (\tau, \bar \tau)\over \sqrt{-g}} =  {R\over 2} + {3\alpha\over 4} \, {\partial \tau \partial \bar \tau\over ({\rm Im}  \tau )^2}- V(\tau, \bar \tau)  \ .
\label{hyper1}\ee
The scalar kinetic term has an unbroken $SL(2,\mathbb{R})$ symmetry, whereas the potential breaks $SL(2,\mathbb{R})$ down to  $SL(2,\mathbb{Z})$.
As we will show, $SL(2,\mathbb{Z})$ symmetry protects the $SL(2,\mathbb{R})$ symmetry of the kinetic term.

$SL(2,\mathbb{R})$  symmetry of the kinetic terms of the cosmological $\alpha$-attractors  was discussed in detail in  \cite{Kallosh:2013yoa,Galante:2014ifa,Kallosh:2015zsa,Ferrara:2013rsa, Carrasco:2015uma,Carrasco:2015rva,Carrasco:2015pla} both in half-plane variables $\tau$ as well as in disk variables $z={\tau-i\over \tau+i}$. 
Their features related to hyperbolic geometry were described in \cite{Achucarro:2017ing,Linde:2018hmx} where it was found that the axion,  the superpartner of the inflaton, tends to freeze during inflation even if the axion potential is not flat. This freezing during inflation at large $\vp$ is due to the kinetic term of the axion in hyperbolic geometry; for example,  $\tau= \theta+i\,e^{-\sqrt{2\over 3\alpha} \varphi}$ in \cite{Linde:2018hmx}.

However, in the $SL(2,\mathbb{Z})$ invariant models studied here, we investigate inflation not close to the boundary $\tau_2\to 0$, as in \cite{Linde:2018hmx}, but in the large  $\tau_2$ limit, and the relevant variable is different, $\tau= \theta+i\,e^{\sqrt{2\over 3\alpha} \varphi}$.  The results of \cite{Achucarro:2017ing,Linde:2018hmx} do not apply here. Therefore we reexamined the behavior of the axion field during inflation. We found that in the $SL(2,\mathbb{Z})$ invariant models studied in this paper, the axion field also does not move during inflation because of the extraordinary double-exponential suppression of the derivatives of the axion potential at large $\vp$.

As a result, just as in the models studied in \cite{Achucarro:2017ing,Linde:2018hmx}, inflationary predictions of the new class of models studied here coincide with the general predictions of single-field $\alpha$-attractors described in Section \ref{old}.  

In Section \ref{SLZ}, we give a brief discussion of $SL(2,\mathbb{Z})$ and $SL(2,\mathbb{R})$ symmetries. In Section~\ref{uni}, we show that $SL(2,\mathbb{Z})$ invariance protects the original $SL(2,\mathbb{R})$ invariance of the kinetic terms, which plays a crucial role in the theory of $\alpha$-attractors.
In Section \ref{Sec:Inv}, we describe various  $SL(2,\mathbb{Z})$  invariants and their properties. In Section \ref{examples}, we propose several $SL(2,\mathbb{Z})$ invariant potentials having an inflationary plateau and study their properties. In Section \ref{Sec: axions}, we give a general outline of inflationary dynamics in these models.
In Section \ref{Sec: twins}, we explore the global structure of $SL(2,\mathbb{Z})$ inflationary potentials and show that they have not one but many flat inflationary plateaus physically equivalent to each other.

The supersymmetric version of the $SL(2,\mathbb{Z})$ symmetric models is given in Section \ref{Sec:super} in the framework of $\overline {D3}$-induced geometric inflation   \cite {Kallosh:2017wnt,McDonough:2016der}   which depends on the superpotential $W(\tau)$, on the \K potential $K(\tau, \bar \tau)$, and on a new real function $G_{X\bar X} (\tau, \bar \tau)$, which is a moduli dependent metric of the nilpotent superfield $X$.
The nilpotent superfield $X^2=0$ represents the upliting effect of $\overline {D3}$ brane in supergravity. A related  construction known as 
 ``liberated supergravity" \cite{Farakos:2018sgq},  also has a potential with   a new arbitrary function ${\cal U} (\tau, \bar \tau)$ in addition to F- and D-terms potentials.  The new function ${\cal U}$ can be related to the metric of the nilpotent superfield 
  under certain conditions so that  $ G_{X\bar X} (\tau, \bar \tau) = e^{K} (\tau, \bar \tau) {\cal U}^{-1} (\tau, \bar \tau)$.  

Our new supersymmetric cosmological models here will be based on \cite {Kallosh:2017wnt}.  By a proper choice of this function one can have easily incorporate any $SL(2,\mathbb{Z})$ invariant potential in supergravity, see Section \ref{Sec:super}.

Predictions of the new $SL(2,\mathbb{Z})$  cosmological models are related to specific discrete targets for the LiteBIRD experiment \cite{LiteBIRD:2022cnt}, known as Poincar\'e disks  \cite{Ferrara:2016fwe,Kallosh:2021vcf}, see Fig. 2 in \cite{LiteBIRD:2022cnt}. These are $\alpha$-attractors with integer values of $3\alpha=1,2,3,4,5,6,7$. The same predictions with integer values of $3\alpha$ are made in M-theory and Type IIB implementations of $\alpha$-attractors \cite{Kallosh:2017ced,Kallosh:2017wnt,Kallosh:2017wku,Gunaydin:2020ric,Kallosh:2021vcf}.
Note that the disk geometry is related to half-plane geometry by a change of variables, $\tau=i {1+z\over 1-z}$, known as the Cayley transform. 

The new $SL(2,\mathbb{Z})$ invariant models based on half-plane geometry, under certain conditions, give the same predictions as the earlier $\a$-attractor models,  including the ones  with Poincar\'e disks. The $SL(2,\mathbb{Z})$ symmetry of the total action has an {\it additional  motivation}. Namely, it was proposed in \cite{Ferrara:1989bc}  that target space modular invariance, which is a property of $SL(2,\mathbb{Z})$ cosmological theories we study here,  is motivated by string theory and by the fact that $SL(2, \mathbb{R})$ symmetry of supergravity can be broken down, due to world-sheet instantons, to its discrete subgroup $SL(2,\mathbb{Z})$. But $SL(2,\mathbb{Z})$ is believed to be preserved  non-perturbatively.

\section{\boldmath Warm up: $\alpha$-attractor models}\label{old}
Consider the following theory
\be\label{acE}
{ {\cal L} \over \sqrt{-g}} =  {R\over 2} - {3\alpha\over 4} \, {(\partial \rho)^2\over \rho^{2}}- V(\rho) \ .
\ee
where the  potentials $V(\rho)$   are non-singular near the pole at $\rho = 0$, and can be expanded in series 
\be\label{EC}
V= V_0 \left(1 - c \rho +{\cal O}(\rho^2)\right)
\ee
with $c > 0$. Since the kinetic term is invariant with respect to the rescaling of $\rho$, one can replace $\rho \to \rho/c$. In these rescaled variables, the potential in the vicinity of the pole acquires the universal form 
\be
V= V_0(1  - \rho +\cdots )
\label{exp}\ee
assuming that $\rho^n$ with $n\geq 2$ can be neglected.
One can  represent ${\cal L}$ in canonical variables $\vp$, where $\rho = e^{-\sqrt {2 \over 3\alpha}\,\vp}$. This yields
\be\label{Lpole}
{ {\cal L} \over \sqrt{-g}} =  {R\over 2} - {1\over 2}  (\partial \varphi)^2- V( e^{-\sqrt {2\over 3\alpha}\varphi}) \ . 
\ee
At large values of $\vp$, such that $e^{-\sqrt{2\over 3\alpha} \varphi } \ll 1$, the potential is
\be\label{expcan}
V= V_0\left(1 -  e^{-\sqrt{2\over 3\alpha} \varphi} + {\cal O}(e^{-2\sqrt{2\over 3\alpha} \varphi})\right) \ .
\ee
During inflation at $\varphi  \gg \sqrt{3\alpha\over 2}$, the potential is determined by the first two terms in the expansion, containing only two parameters, $V_{0}$ and $\alpha$. In this approximation, one can calculate the total expansion of the universe $e^{N_{e}}$ during inflation when the field rolls down from $\vp = \vp_{N_{e}}$:
\be
 e^{-\sqrt{2\over 3\alpha} \varphi_{_{N_e}}} = {3\alpha\over 8N_{e}}. 
\label{efold}\ee 
Therefore, the amplitude of inflationary perturbations generated at large values of the field $\vp$ corresponding to $N_{e} \sim 55$ and $\alpha \lesssim O(1)$  does not depend on higher order terms in the expansion \rf{expcan}. In particular, the amplitude of inflationary perturbations $A_{s}$,  the spectral index $n_{s}$, and the tensor to scalar ratio $r$ in the large $N_{e}$ limit are given by \cite{Kallosh:2013yoa,Galante:2014ifa,Kallosh:2021mnu}
\be
\label{pred}
 A_{s} = {V_{0}\, N_{e}^{2}\over 18 \pi^{2 }\alpha} \ , \qquad n_{s} = 1-{2\over N_{e}} \ , \qquad r = {12\alpha\over N^{2}_{e}} \ .
\ee 
These predictions depend only on two parameters, $V_{0}$ and $\alpha$, and do not depend on many other features of the original potential $V(\rho)$. That is why we called these models  $\alpha$-attractors. 

Some of the simplest $\alpha$-attractor potentials  are E-models 
\be\label{EEpot}
V= V_0 \left(1 -\rho\right)^{2n} =  V_{0}\left(1-  e^{-\sqrt{2\over 3\alpha} \varphi }\right)^{2n} \  .
\ee
These potentials have an inflationary plateau at $\vp \gg \sqrt{3\alpha\over 2}$, and they blow up at large negative values of $\vp$.
For $n = 1$ and $\alpha = 1$, the potential \rf{EEpot} coincides with the potential of the Starobinsky model. In general, $\alpha$ can take any value, which allows to describe observational data with any $r$, all the way down to zero. On the other hand, in models based on maximal supergravity, $3\alpha$ may take 7 integer values, $3\alpha = 1, 2,..., 7$ \cite{Ferrara:2016fwe,Kallosh:2021vcf}.
\begin{figure}[H]
\centering
\includegraphics[scale=0.41]{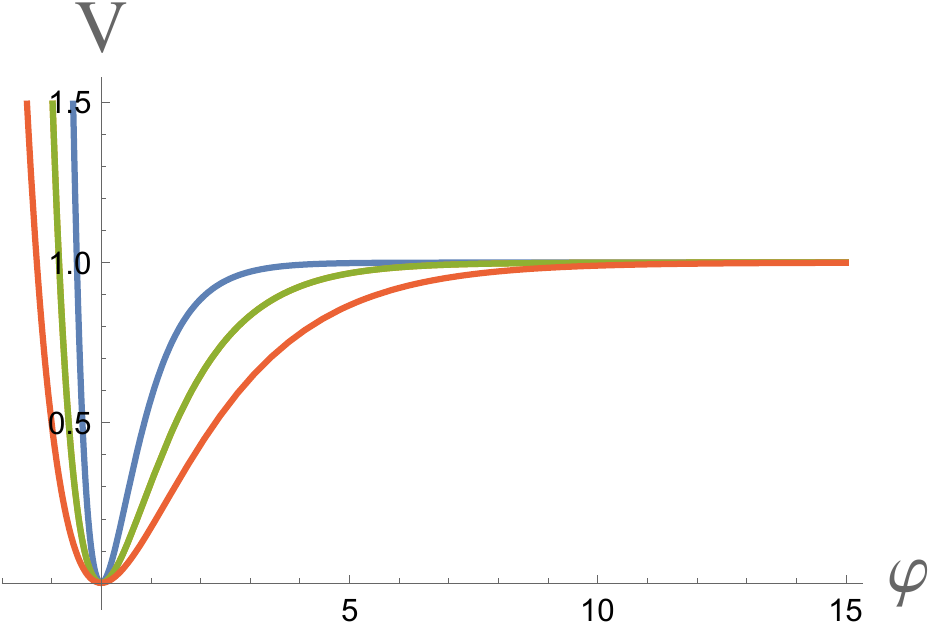}
\vskip -10pt
\caption{\footnotesize E-model  potential \rf{EEpot} for $V_{0} = 1$, $n=1$, $3\alpha = 1$ (blue line), $3\alpha = 3$ (green line, Starobinsky model), and $3\alpha = 7$ (red line).   The potential blows up at large negative values of $\vp$ and has a plateau at large positive $\vp$.}
\label{EEF}
\end{figure}

Ref. \cite{Casas:2024jbw} suggested that the $SL(2,\mathbb{Z})$ invariant inflationary models are similar to the Starobinsky model, but the authors noted that this similarity is incomplete. Indeed, the potential of the Starobinsky model, as well as the potentials of  E-models, break inversion symmetry, and therefore they are not $SL(2,\mathbb{Z})$ invariant. 

As we will see, $SL(2,\mathbb{Z})$  cosmological $\alpha$-attractors are more closely related to another family of $\alpha$-attractors, T-models. The simplest T-model 
has the potential 
 \be\label{TT}
V = V_{0}\left({1- \rho  \over  1+ \rho}\right)^{2n}  = V_{0} \tanh^{2n} {\varphi\over \sqrt{6\alpha} }  \ .
 \ee 
 This  potential has the  expansion 
$V= V_0(1 - 4n  \rho +{\cal O}(\rho^2))$ 
 near the pole at $\rho = 0$.   Thus, this model belongs to the general class of $\alpha$-attractors and has the same cosmological predictions \rf{pred} for $N_{e}\gg 1$ and $\alpha \lesssim O(5)$.   General  T-models potentials  can be represented as  $V =F\left(\tanh^{2} {\varphi\over \sqrt{6\alpha} }\right)$ \cite{Kallosh:2013yoa,Galante:2014ifa,Kallosh:2015zsa}.

\begin{figure}[H]
\centering
\includegraphics[scale=0.38]{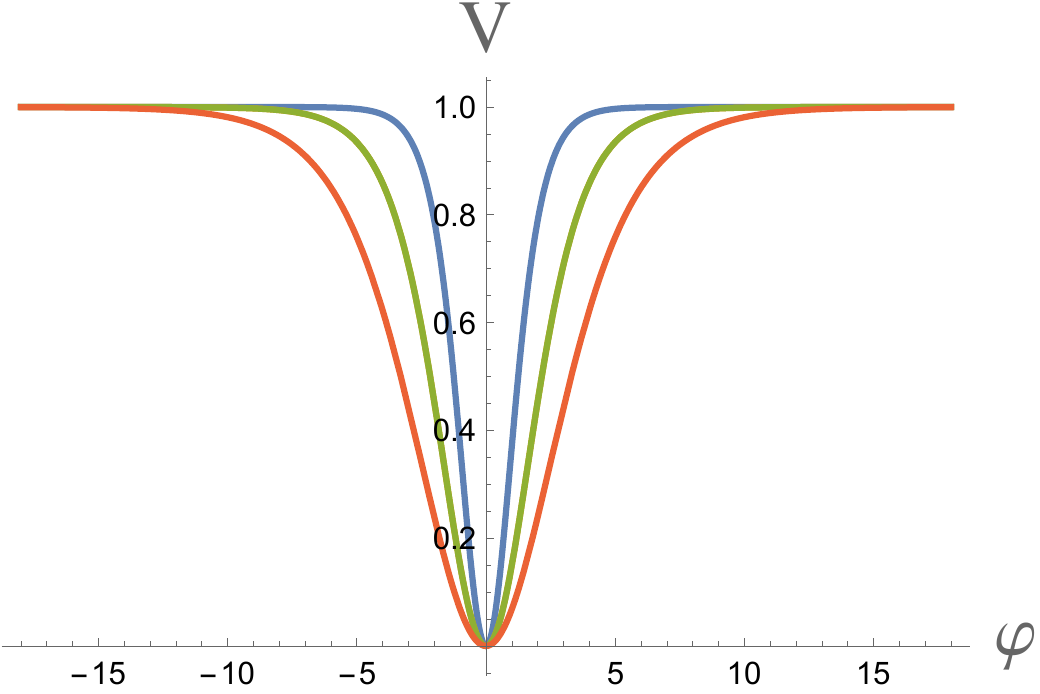}
\vskip -10pt
\caption{\footnotesize T-model  potential \rf{TT} for $V_{0} = 1$, $3\alpha = 1$ (blue line), $3\alpha = 3$ (green line), and $3\alpha = 7$ (red line).   }
\label{TTF}
\end{figure}

Unlike the E-models and the Starobinsky model, the theory \rf{acE} with the T-model potential \rf{TT} has the inversion symmetry 
\be
\rho \to \rho^{-1}\, , \qquad \varphi \to -\varphi  \ .
\ee
Therefore, the potentials of T-models  \rf{TT} look like the letter T: They have two symmetric plateaus, shoulders of the same height, at  $|\vp| \gg \sqrt{3\alpha\over 2}$ \cite{Kallosh:2013hoa,Kallosh:2013yoa}.  We show these potentials for $V_{0} = 1$ and $3\alpha = 1,  3, 7$ in Fig. \ref{TTF}.

\begin{figure}[H]
\centering
\includegraphics[scale=0.26]{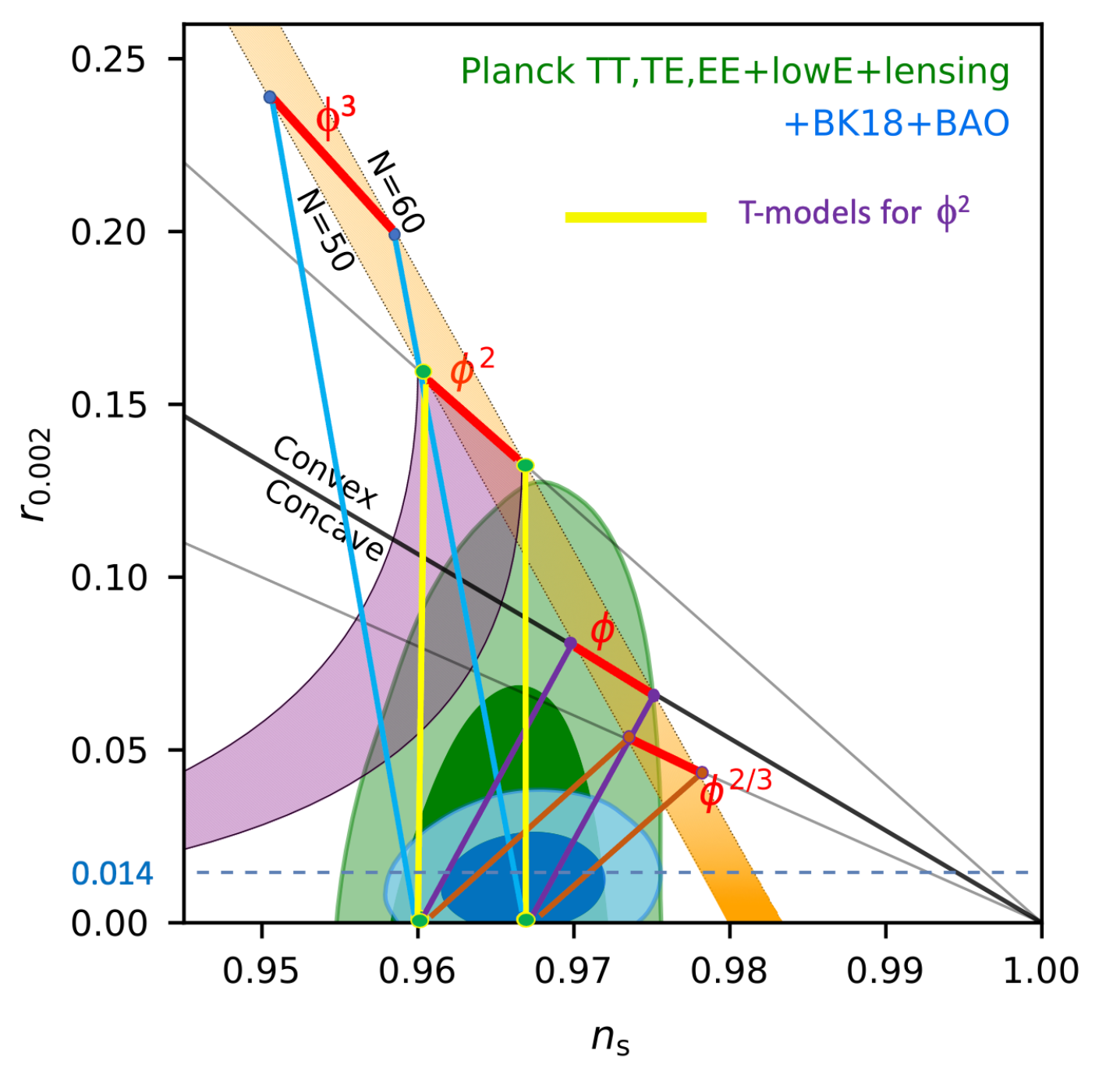}
\vskip -10pt
\caption{\footnotesize The figure shows the main results of the BICEP/Keck \cite{BICEPKeck:2021gln} superimposed with the predictions of  $\alpha$-attractor T-models with the potential $\tanh^{2n} {\varphi\over \sqrt{6\alpha}}$ \cite{Kallosh:2013yoa,Kallosh:2015zsa,Kallosh:2021mnu}. At $\alpha \gg 1$,  these models have different predictions corresponding to different potentials  $\phi^{n}$ (at $\alpha \rightarrow \infty$), but with a decrease of $\alpha$, these predictions converge in the dark blue area favored by CMB observations. }
\label{Fan}
\end{figure}

Cosmological predictions of these models are compatible with all presently available CMB-related observational data for  $\alpha \lesssim O(5)$ \cite{Kallosh:2021mnu}, see Fig. \ref{Fan}.

\section{Symmetries: $SL(2,\mathbb{R})$ and $SL(2,\mathbb{Z})$}\label{SLZ}

Here we study the action in eq. \rf{hyper1} which depends on modulus $\tau(x)$ where the field $\tau$  is defined  in  Poincar\'e {\it upper half-plane} such that
\be
{\cal H}= \{ \tau\in \mathbb{C}, {\rm Im} \, \tau >0 \}
\ee
and  $\tau= \tau_1+i \tau_2$ is a complex modulus. A kinetic term for scalar fields $\tau$ 
 in eq. \rf{hyper1} of the form $ {\partial \tau \partial \bar \tau\over ({\rm Im}  \tau )^2}$, has an $SL(2,\mathbb{R})$ symmetry. The action of $SL(2,\mathbb{R})$ maps the upper half-plane  to itself and 
 \be
 (\Gamma \cdot \tau) \equiv \frac{a\tau + b}{c \tau + d}, \quad {\rm where}\,\, \Gamma=\begin{pmatrix}
a & b \\
c & d
\end{pmatrix}\, , \quad \det \Gamma=ad-cb=1 \ .
\label{sym}\ee  
All parameters  $a,b,c,d$  of  $SL(2,\mathbb{R})$ symmetry are real numbers.
The symmetries can be described  as follows: 
\begin{itemize}
  \item Translation:\ \           $\tau \to \tau +b/d$   , \ $\Gamma_{transl}= \begin{pmatrix}
d & b \\
0 & d
\end{pmatrix}$
  \item Dilatation of the entire plane: \ \  \, $\tau \to a^2 \tau$   ,  \ $\Gamma_{dil}= \begin{pmatrix}
a & 0 \\
0 & 1/a
\end{pmatrix}$
  \item Inversion: \ \  $\tau \to -{1\over c^2 \tau}$   ,  \  $\Gamma_{inv}= \begin{pmatrix}
0 & -1/c \\
c & 0
\end{pmatrix}$ 
\end{itemize}
The full symmetry of the kinetic term in \rf{hyper1} was described in \cite{Carrasco:2015uma} where it was noticed that in addition to M\"obius symmetry \rf{sym} where $\det \Gamma=ad-cb\neq 0$ one also have a reflection of $\tau_1$, i.e. of the real part of $\tau$:
\begin{itemize}
\item Reflection
$
\tau_1\to -\tau_1
$
\end{itemize} 
The $SL(2,\mathbb{Z})$ symmetry\footnote{More general discrete congruence subgroups of the M\"obius $SL(2,\mathbb{R})$ group are
Hecke subgroups $\Gamma_0(N)$ of level $N$, defined by matrices in eq. \rf{sym} where $c$ is not an arbitrary integer, but a multiple of $N$.}
has only integer parameters and is usually described by the action of two generators 
\begin{itemize}
  \item Translation: \ \    $\tau \to \tau +1$\,  , $\Gamma_{transl}= \begin{pmatrix}
1 & 1 \\
0 & 1
\end{pmatrix}$
  \item Inversion:  \ \ $\tau \to -{1\over  \tau}$\,  ,   $\Gamma_{inv}= \begin{pmatrix}
0 & -1 \\
1& 0
\end{pmatrix}$   
\end{itemize} 
But reflection is also a property of the kinetic term in  \rf{hyper1} since it is an isometry of the geometry 
 \begin{itemize}
\item Reflection
$
\tau_1\to -\tau_1
$
\end{itemize} 

Note that the dilatation symmetry is not present anymore since in the matrix $\Gamma_{dil}= \begin{pmatrix}
a & 0 \\
0 & 1/a
\end{pmatrix}$
both $a$ and $1/a$ must be integers. Therefore only a trivial dilatation is possible, $\tau \to a^2 \tau$ where $a^2=1$.

There are two special (self-dual) points
points for $SL(2,\mathbb{Z})$ symmetry with $|\tau |=1$
\be
 |\tau |=1 : \qquad \tau=i \, \,  \, (  \tau_1=0, \tau_2=1)\ee and 
\be
\, \, \, \, \, |\tau |=1:  \qquad \tau=e^{{2\pi i\over 3} }= - {1\over 2} + i \sqrt{3/4}\ee 
We will find out that our $SL(2,\mathbb{Z})$ invariant potentials have a saddle point at $\tau =i$ and a minimum at $\tau=e^{{2\pi i\over 3} }$.

\section{Kinetic term universality}\label{uni}
A standard $SL(2,\mathbb{R})$ invariant supergravity kinetic term for $\tau =i T$ is
\be-{3\alpha\over 4} \, {\partial \tau \partial \bar \tau\over ({\rm Im}  \tau )^2}= -{3\alpha\over 4} \, {\partial T \partial \bar T\over ({\rm Re}  T )^2}\ee
 follows from the \K potential $K= -3\alpha \ln (T+\bar T))$. There are corrections to \K\, potentials due to various string theory corrections,  which were discussed in the literature. In the case of one modulus, the detailed study was performed in  \cite{Broy:2015qna}. In the case of both the string 1-loop and the leading ${\cal O}(\alpha')^3$-corrections to the type IIB volume moduli \K potential the study of the effect of these corrections on the pole inflation in $\alpha=1$ case is
\be
K+\delta K= -3 \ln (T+\bar T)- {\xi\over (T+\bar T)^{3/2}}- {C\over (T+\bar T)} - {D\over (T+\bar T)^2} \ ,
\ee
and we define
\be
\delta K= \delta K_{3/2}+ \delta K_{1}+ \delta K_{2} \ .
\ee
We would like to find out how these additional terms affect the kinetic term. We compute the 
\K metric to find
\be
\partial_T \partial_{\bar T}  \delta K_l  = l(l+1) { \delta K_l \over (T +\bar T)^{2}} \ .
\ee
This means that the kinetic terms of the form 
\be
 {\partial T \partial \bar T \over (T +\bar T)^{7/2}}, \, \qquad { \partial T \partial \bar T \over (T +\bar T)^{3}}\, , \qquad { \partial T \partial \bar T \over (T +\bar T)^{4}}
\ee
are added to the original kinetic term  of the form ${ \partial T \partial \bar T \over (T +\bar T)^{2}}
$. Corrections to kinetic term preserve $SL(2,\mathbb{Z})$  translation symmetry $T\to T-i$, however, they break inversion $T \to {1\over T}$. We conclude that the requirement of $SL(2,\mathbb{Z})$ inversion excludes these corrections.

This observation is consistent with the fact that in supersymmetric theories, quantum corrections do not involve terms like $R^2$ and $R^3$, and gauge-invariant corrections usually start with terms like $R^4$ and higher. These should not affect the kinetic terms but might affect the potentials.

Thus, we conclude that the requirement of $SL(2,\mathbb{Z})$ symmetry of the kinetic term preserves the classical  \K potential for one modulus. It is interesting that by requiring $SL(2,\mathbb{Z})$ symmetry of the kinetic term we have actually preserved the classical kinetic term which has an unbroken $SL(2,\mathbb{R})$ symmetry. Namely, our $SL(2,\mathbb{Z})$-invariant  kinetic term in eq. \rf{hyper1}  has a bonus symmetry, a continuous dilatation symmetry
\be
T\to a^2 T\, , \qquad a\in \mathbb{R} \ ,
\label{bonus}\ee
since it is $SL(2,\mathbb{R})$ and not just $SL(2,\mathbb{Z})$ invariant. This bonus symmetry will play an important role in the attractor properties of our new cosmological models where the potential has an exact $SL(2,\mathbb{Z})$ symmetry without dilatation present in $SL(2,\mathbb{R})$ symmetry.

\section{$SL(2,\mathbb{Z})$ invariants }\label{Sec:Inv}
We may represent $\tau $ as  
$\tau_1+\tau_2= \tau_1+ie^{\sqrt{2\over 3\alpha} \varphi}$.

1. One  $SL(2,\mathbb{Z})$ invariant can be given is the form depending on  $ j$-invariant\footnote{The $j$-inflation models, where potentials depend in $j$-invariant, were proposed and studied in  \cite{Schimmrigk:2014ica,Schimmrigk:2016bde,Schimmrigk:2021tlv}.  The choice of potentials in $j$-inflation models was 
 $V\sim |j(\tau)| ^{2p}$.  The difference with our potentials in Sec. \ref{Sec:j} is that at large $\tau_2$ $j$-inflation potentials behave as $e^{4p\pi \tau_2} $ whereas our $j$-invariant potentials approach a plateau.
}
\be\label{Klein}
j(\tau)= 12^3 J(\tau) \ ,
\ee
where $J(\tau)$ is known as Felix's Klein Absolute Invariant with the properties $J(i)=1$, $J(e^{2\pi i/3})=0$.
\be
j(\tau) = q^{-1} + \sum_{n=0} c_n q^n \, , \qquad  q=e^{2\pi i  \tau} \ .
\label{jfunc}\ee
At $\tau=i$ this function is $j(i)=12^3$,  and at $\tau=e^{i\pi\over 3}= {1\over 2}(- 1+i \sqrt 3)=0$ it vanishes,  
\be
j(i)=12^3 \ , \qquad j(e^{i\pi\over 3})=0\, . 
\ee
It  is negative at the points $\tau = {1\over 2} (1+ i \sqrt {-H})$, for  the Hegner numbers 
$$
H= -7, -11, -11, -19, -43, -67, -163
$$ 
 We note that   the function $|j(\tau)|^2 \equiv   j(\tau) \bar j(\bar\tau) \geq 0$ and define a 
 ``shifted''  function so that it has no zeros at $\tau = e^{i\pi\over 3}$:  
\be
| j(\tau)|^2_{shift}=| j(\tau)|^2 + \b^2\label{b}\ee 
where $\b$ is some constant. For $\beta > 1$ we have  $| j(\tau)|^2_{shift}>1$ and therefore its logarithm is positive. 

At large $\tau_2$    the sum in \rf{jfunc} is dominated by the first term, and therefore
\be
|j(\tau)|^2 \equiv   j(\tau) \bar j(\bar\tau) \to  e^{4\pi \tau_2} \ .
\label{large}\ee
where  in the leading approximation one can neglect $\tau_1$ in  $\tau = \tau_1+i\tau_2$.

We define the function 
\be
L_{j}(\tau, \bar \tau) \equiv {1\over 4\pi} \ln ( | j(\tau))|^2 + \beta^2)  > 0 \ .
\label{I3}\ee
 At large $\tau_2$
\be
L_j(\tau, \bar \tau)|_{\tau_2\to \infty} \to  \tau_2 \ .
\label{jlarge}\ee
 Note that in the limit  ${\rm Im}(\tau)  = e^{\sqrt{2\over 3\alpha} \varphi}\to \infty
$, $\vp\to+\infty$ this function does not depend on axions. Meanwhile in the limit  ${\rm Im}(\tau) \to 0$, $\vp\to -\infty$ there is a significant dependence on axions $\tau_1$. We will see in the plots of $SL(2,\mathbb{Z})$ invariant inflationary potentials that at large positive $\vp$ axions form a flat direction, however, at large negative $\vp$, $\tau_2\to 0$ there is a significant dependence on axions as one can see from \rf{jfunc}.

2. Here we use  
the    automorphic form, known as   a regularized non-holomorphic Eisenstein 
series of order  1, studied in the cosmological context  in \cite{Casas:2024jbw} 
\be 
\hat E_1 (\tau, \bar \tau)\equiv 		- \pi  \,  \ln({\rm Im}(\tau) |\eta^2(\tau)|^2) \ .
\label{hatE}	\ee
 where $\eta(\tau)$  is the Dedekind function 
\be
\eta(\tau) = q^{1/24} \prod _{n=1}^{\infty} (1- q^n)\label{eta}\, , \qquad  q=e^{2\pi i  \tau} \ .
\ee 
This function has the properties
 \be
 \eta(\tau+1) = e^{{i\pi\over 12} \tau} \, \eta(\tau)\, \qquad \eta \left (-{1\over \tau}\right)= \sqrt {-i \tau} \, \eta(\tau) \ .
 \ee
 Thus ${\rm Im}(\tau)$ and $\eta(\tau) \bar \eta(\bar\tau)$ are  invariant under $\tau \to \tau+1$. Under inversion $\tau \to -{1\over  \tau}$
 \be
{\rm Im \tau } \to   {1\over \tau \bar \tau }  {\rm Im \tau }\, \qquad |\eta^2(\tau)|^2 \to \tau \bar \tau |\eta^2(\tau)|^2 \ .
 \ee
 An important property of the Dedekind function  derived in \cite{Atiyah1987}  is
 \be
   \ln \eta (\tau)  \to  -{ \pi\over 12}  {\rm Im}(\tau) +{\cal O} ( e^{2\pi i  \tau})  \, \qquad {\rm as} \qquad {\rm Im}(\tau) \to \infty \ .
\label{Atiyah} \ee
Here again at  ${\rm Im}(\tau)  = e^{\sqrt{2\over 3\alpha} \varphi}\to \infty
$, $\vp\to \infty$  the limit does not depend on axions, but a limit to  ${\rm Im}(\tau) \to 0$, $\vp\to -\infty$ depends on axions significantly.   Thus, these flat axion directions during inflation, which we will see in all our examples, follow directly from the result in  
\cite{Atiyah1987}.

We define 
\be 
L_{\eta}(\tau, \bar \tau)\equiv {3\over \pi^2}  \hat E_1 (\tau, \bar \tau)= - {3\over \pi}   \,  \ln ( {\rm Im}(\tau) |\eta^2(\tau)|^2) ) = 	 - {3\over \pi}   \,  \ln ( |\eta^2(\tau)|^2) - {3\over \pi}   \,  \ln({\rm Im}(\tau))\
\label{I1}	\ee
Thus,  according to  \cite{Atiyah1987}  as shown in eq. \rf{Atiyah} at $ {\rm Im}(\tau) \to \infty$
 \be 
L_{\eta}(\tau, \bar \tau)\equiv {3\over \pi^2}  \hat E_1 (\tau, \bar \tau) \to   \tau_2 -  {3\over \pi} \,  \ln( \tau_2) + {3\over \pi^2}{\cal O} (e^{-2\pi \tau_2}) \ .
\label{limit}	\ee
 In the past the $SL(2,\mathbb{Z})$ invariant $ -\ln({\rm Im}(\tau) |\eta^2(\tau)|^2)$ was derived in \cite{Dixon:1990pc} in the computations of the moduli dependence of string loop corrections to gauge coupling constants where the moduli dependent correction to ${1\over g^2}$  was proportional to $\hat E_1 (\tau, \bar \tau) $. 
 More recently the same 
$SL(2,\mathbb{Z})$ invariant was studied in \cite{Green:2010wi} where it was related to the presence of a logarithmic singularity in the one-loop graviton scattering amplitude in 8-dimensional supergravity. 

At large $\tau_2$ it is safe to ignore in \rf{limit} the terms ${3\over \pi^2}{\cal O} (e^{-2\pi \tau_2})
$, however, the second term requires attention.
\be 
{3\over \pi^2} \hat E_1 (\tau, \bar \tau)|_{\tau_2 \to \infty, \tau_1=0}	 \simeq  \tau_2\left (1  -  {3\over  \pi} \,  {\ln( \tau_2)\over \tau_2} \right) \ .
\label{as}	\ee	
Thus the deviation from the simple T-models involves the factor 
\be
{3\over  \pi} \,  {\ln( \tau_2)\over \tau_2} \simeq  \Big (\varphi_{_{N_e}} \sqrt{2\over 3\alpha} \Big) e^{-\sqrt{2\over 3\alpha} \varphi_{_{N_e}}}   \ .
\label{ev}\ee	
Using \rf{efold} we find that for example, at $N_e=55$ and $3\alpha=1$  the correction term in eq. \rf{as} during inflation is still quite small \footnote{In \cite{Linde:2018hmx} the term ${1\over \tau_2} =e^{-\sqrt{2\over 3\alpha} \varphi_{_{N_e}}}$ was evaluated during inflation and it was found to be $2\times 10^{-3}$.}, even in presence of a factor $ \ln( \tau_2)$.
\be
{3\over  \pi} \,  {\ln( \tau_2)\over \tau_2} \approx 10^{-2} \ .
\label{ev1}\ee	

3. Another $SL(2,\mathbb{Z})$ invariant used in \cite{Casas:2024jbw} involves an almost holomorphic modular form of weight 2 known as $\tilde G_2= G_2-{\pi\over \tau_2}$, where $G_2=-4\pi i \partial_\tau \ln \eta(\tau)$
\be
L_{G_2}(\tau, \bar \tau)\equiv {3^2\over \pi^2} |\tau_2 \tilde G_2|^2 = {3^2\over \pi^2}\left ({\tau_2 G_2\over \pi} -1\right) \left ({\tau_2 \bar G_2\over \pi} -1\right) \ .
\ee
At $ {\rm Im}(\tau) = \tau_2\to \infty$
\be
L_{G_2}(\tau, \bar \tau) \to  {3^2\over \pi^2} \left(1-{\tau_2}{\pi\over 3}\right)^2 \to  {3^2\over \pi^2}\tau_2^2  \Big ({\pi\over 3} - {1\over \tau_2}\Big )^2 
\ee
since $G_2\to {\pi^2\over 3}$ at $ {\rm Im}(\tau) \to \infty$.

Thus we propose to use the action which has an $SL(2,\mathbb{Z})$ symmetry.  We suggest to look for potentials depending on $SL(2,\mathbb{Z})$ invariants
\be\boxed{
\left( {\cal L} (\tau, \bar \tau)\over \sqrt{-g}\right)^{SL(2, \mathbb{Z})} =  {R\over 2} + {3\alpha\over 4} \, {\partial \tau \partial \bar \tau\over ({\rm Im}  \tau )^2}- V\Big (L_j (\tau, \bar \tau), L_{\eta}(\tau, \bar \tau), L_{G_2}(\tau, \bar \tau), \dots  \Big ) }
\label{hyper3}\ee
Here $ \dots$ include other $SL(2,\mathbb{Z})$ invariants, which we have not discussed here.

 If the  candidate for potentials  $V\Big (L_j (\tau, \bar \tau), L_{\eta}(\tau, \bar \tau), L_{G_2}(\tau, \bar \tau), \dots  \Big )$ have the expansion at $ {\rm Im}(\tau) \to \infty$ of the form 
 $V_0(1-{c\over {\rm Im}(\tau)})$, neglecting terms ${\cal O} ({1\over {\rm Im}(\tau)})^2$, the dependence on $c$ will be absorbed into a redefinition of the kinetic term, thanks to bonus symmetry \rf{bonus}. Therefore the theory in the inflationary regime will depend only on $\alpha$ and $V_0$, as in the case of the pole inflation \cite{Galante:2014ifa} with $\rho\sim {1\over {\rm Im}(\tau)}$, and as explained in detail in Sec. \ref{old}.

 We will give examples below of the general formula in eq. \rf{hyper3}. Clearly, more of the $SL(2,\mathbb{Z})$ invariant potentials can be constructed, which will be the same near the attractor point.

\section{Examples of $SL(2,\mathbb{Z})$ invariant potentials}\label{examples}

\subsection{Examples with  $ j$-function}\label{Sec:j}
We will introduce two potentials, 
\be\label{Renata1}
V_{1}^{n}(\tau, \bar \tau)= V_0 \Big ({I(\tau, \bar \tau) - 1\over I(\tau, \bar \tau) + 1}\Big)^n \ ,
\ee
\be\label{Renata2}
V_{2}^{n}(\tau, \bar \tau)= V_0 \Big (1-{I(\tau, \bar \tau)^{-1} }\Big)^n \ ,
\ee
where
\be
I (\tau, \bar \tau)\equiv {\ln ( | j(\tau))|^2 + \beta^2) 
\over \ln \beta^2} \  .
\ee
 Here the function $I (\tau, \bar \tau)$  is proportional to the function  $ L_j(\tau, \bar \tau)  \equiv {1\over 4\pi} \ln ( | j(\tau))|^2 + \beta^2)$ introduced earlier in eq. \rf{I3}.  
  
We now   make a choice  $\beta= j(i)$ where $j(i)=12^3$.   We made this choice because the function $I (\tau, \bar \tau)$ can be defined in terms of Klein's absolute invariant  $J(\tau)$, where
$
j (\tau)= 12^3 J (\tau) 
$.
Our choice of $\beta= j(i)$ corresponds to the special point $\tau = i$, where $J(i)=1$:
\be
J(i)=1 \qquad \Rightarrow \qquad j (i)=12^3 \ .
\ee
In this case, 
\be
I (\tau, \bar \tau)|_{\beta= j(i)} \equiv {\ln ( | j(\tau))|^2 +  j^{2}(i)) 
\over \ln |j^2(i)|} \ .
\label{J1}\ee
Our function \rf{J1} at $\tau=i, e^{2\pi i/3}$ has the properties
\be
I (\tau, \bar \tau)|_{\tau=i}=  {\ln | 2j(i))|\over \ln |j(i)|} >1\, , \qquad I (\tau, \bar \tau)|_{\tau=e^{2\pi i/3}}=  {\ln | j(i))|\over \ln |j(i)|} =1 \ .
\ee
The derivatives vanish at these two points corresponding to the extrema of the potentials $V_{k}^n$, $k=1,2$. 
\be
\partial_\tau I (\tau, \bar \tau)|_{\tau=i, \tau=e^{2\pi i/3}}=0 \ , \qquad \partial_\tau V_k^n (\tau, \bar \tau)|_{\tau=i, \tau=e^{2\pi i/3}}=0 \ .
\ee
Both potentials  are positive at $\tau=i$,
\be
V_k (\tau, \bar \tau)|_{\tau=i} >0 \ ,
\ee
and  vanish at $\tau=e^{2\pi i/3}$,
\be
V_k (\tau, \bar \tau)|_{\tau=e^{2\pi i/3}} =0 \ .
\ee
As one can also see from the plots, these potentials have a de Sitter saddle point at $\tau=i$ and a minimum at $\tau=e^{2\pi i/3}$ with $V_k = 0$.

We should note that one can add any constant $\Lambda$ to the potentials without breaking their $SL(2,\mathbb{Z})$ invariance. For $\Lambda> 0$, the minimum at $\tau=e^{2\pi i/3}$ would correspond to a dS state with $V_k = \Lambda$.

\begin{figure}[H]
\centering
\includegraphics[scale=0.36]{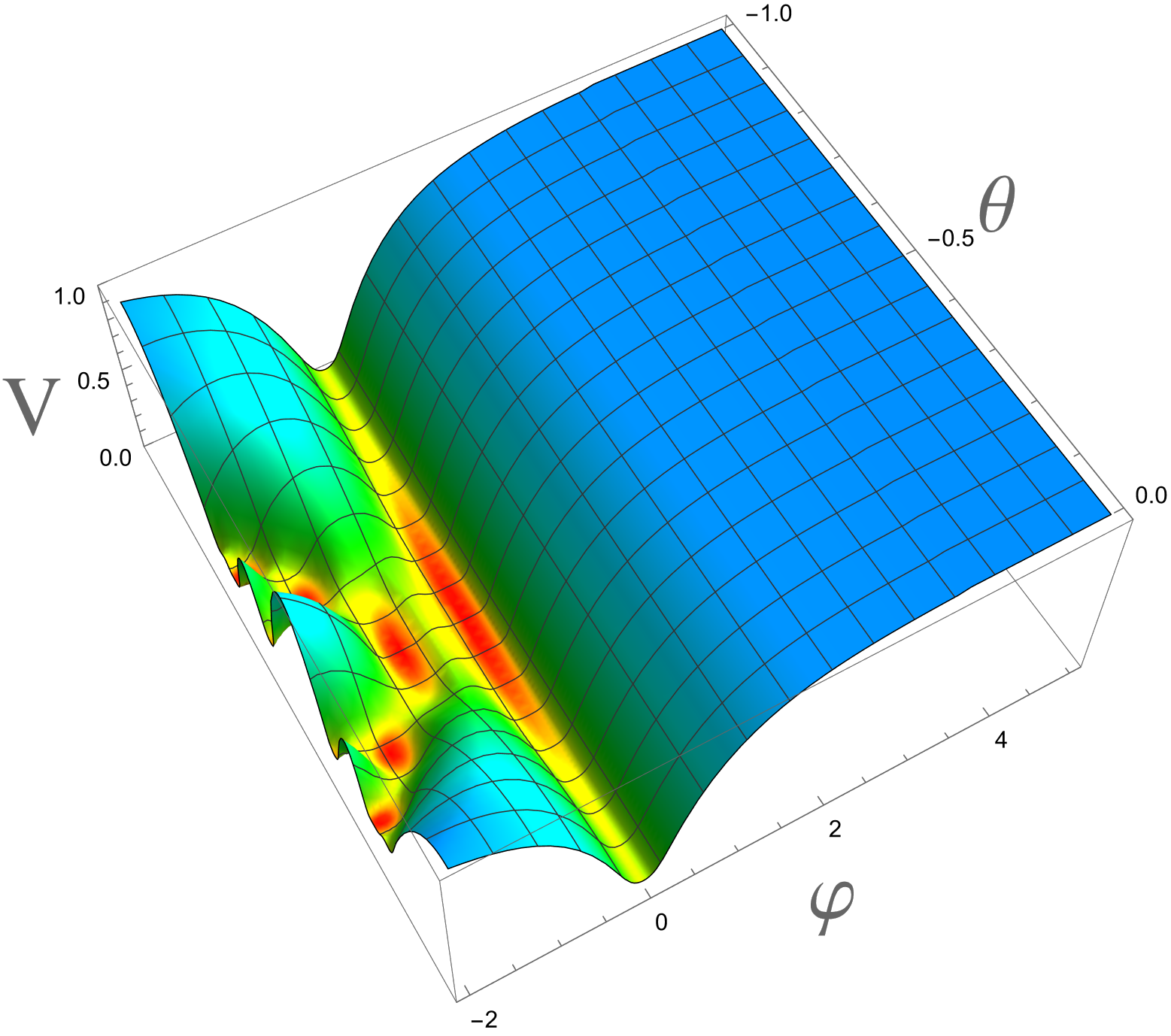}
\vskip -10pt
\caption{\footnotesize  Potential \rf{Vbeta2}  for $n = 1$, $V_{0} = 1$, $\alpha=1/3$, $\beta= j(i)$. The potential is positive. The height of the potential is color-coded, from blue to red. The blue plateau approaches $V_{0}= 1$ at $\vp \to +\infty$. Red spots correspond to $V \ll 1$, which helps to visually identify the minima of the potential. All minima have the same depth $V = 0$, but one can uplift all of them by adding a tiny constant $\Lambda$ to the potential. Potential \rf{Vbeta1} has a very similar shape.}
\label{Renata1fig}
\end{figure}

\begin{figure}[H]
\centering
\includegraphics[scale=0.37]{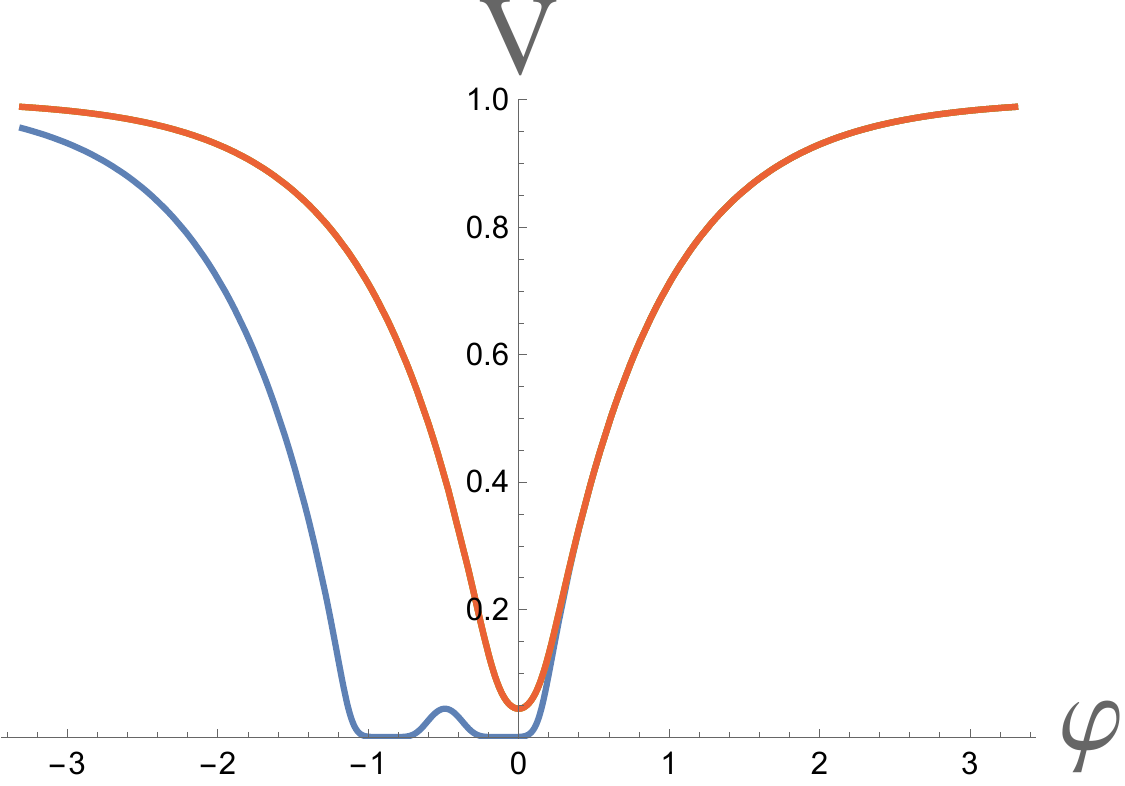}
\vskip -10pt
\caption{\footnotesize 2D slice of the potential \rf{Vbeta2}  shown in Fig. \ref{Renata1fig}  at $\tau_1=\theta=-0.5$ (blue curve), and at $\tau_1=\theta=0$ (bed curve). The red curve coincides with the blue one at large values of $\vp$, which shows that the inflationary potential at large $\vp$ practically does not depend on $\theta$. The unstable de Sitter saddle point can be seen here at $\tau=i$, i.e., at $ \vp=0, \theta=0$,  in the red curve at $V=4\times 10^{-2}$. The blue curve shows that at $\theta = -0.5$, the potential has two minima of equal depth, the first of which is at  $e^{{\sqrt 2} \vp}= {\sqrt 3\over 2}$, the second one at $e^{{\sqrt 2} \vp}={1\over 2}$. One can also see these two minima in Fig. \ref{Renata1fig} as two deep red areas at $\theta = -0.5$.}
\label{2D}
\end{figure}

 At large $\tau_2$, according to \rf{large},
$
|j(\tau)|^2  \to  e^{4\pi \tau_2}$.
 Therefore
\be
I (\tau, \bar \tau)|_{\tau_2\to \infty} = {\ln | e^{2\pi \tau_2})|^2\over \ln |j(i)|^2}= {4\pi \over \ln |j(i)|^2} \tau_2\equiv c\tau_2 \ .
\ee
\be
V_T^n (\tau, \bar \tau)|_{\tau_2 \to \infty} = V_0 \Big ({c\tau_2 - 1\over c\tau_2 + 1}\Big)^n \approx V_0 (1-{4n\over c}{1\over \tau_2} +\cdots) \ ,
\ee
and 
\be
V_E^n (\tau, \bar \tau)|_{\tau_2 \to \infty} = V_0 \Big (1 - {1\over c\tau_2} \Big)^n \approx V_0 (1-{2n\over c}{1\over \tau_2} +\cdots) 
\ee
In the large $\tau_2$ limit, both potentials have a plateau behavior. To study inflation in these models, we use the variables  $\tau_1=\theta$, $\tau_2= e^{\sqrt{2\over 3 \alpha} \varphi}$ where $\vp$ is a canonically normalized inflaton field. The potentials \rf{Renata1},\rf{Renata2} look as follows:
\be\label{Vbeta1}
V_1^{n}  = V_{0}\left({\ln \big (|j(\theta +i e^{\sqrt{2\over 3\alpha}\vp})|^{2} + \beta^{2}\big)     - \ln \beta^{2}
\over  \ln \big(|j(\theta +i e^{\sqrt{2\over 3\alpha}\vp})|^{2} + \beta^{2}\big)    +\ln \beta^{2}}\right)^{n}  \ ,
\ee
 \be\label{Vbeta2}
V_2^{n}  =V_{0}\left(1-{\ln \beta^{2}
\over  \ln \big(|j(\theta +i e^{\sqrt{2\over 3\alpha}\vp})|^{2} + \beta^{2}\big)}\right)^{n} \ .
\ee
In Fig. \ref{Renata1fig} we show the potential $V_{1}^{n}$ for $n = 1$, $\beta =  j(i)$, $\alpha = 1/3$. In Fig. \ref{2D} we show slices of this potential at $\theta=0$ and at $\theta = -0.5$. By looking at these 3D and 2D  potentials, one can see that the cosmological evolution, starting at the plateau at large positive $\vp$,  proceeds as follows. For a while, the axion $\theta$ stays close to its initial value, whereas the inflaton $\vp$ rolls down towards its smaller values. At small $\vp$,  inflation ends, and both fields $\vp$ and $\theta$ move towards  the minima with $V = 0$ indicated by red in Fig. \ref{Renata1fig}  (e.g. the nearest to the plateau is the minimum at $\tau=e^{2\pi i/3}= -{1\over 2} +i {\sqrt 3\over 2}$ and the second one at smaller $\tau_2$ is at $\tau =-{1\over 2} +i {1\over 2}$ ). A more detailed discussion of inflation in this model is contained in Section \ref{Sec: axions}.

In most of the equations in this section, and in the potential \rf{Vbeta2} shown in Fig. \ref{Renata1fig}, we made a particular choice of the parameter $\beta$ introduced in Eq. \rf{I3}: $\beta = j(i) = 12^{3}$. As we already mentioned, this is a natural choice related to the specific normalization of the function $j(\tau) = 12^{3} J(\tau)$ where $J(\tau)$ is the Felix's Klein Absolute Invariant, $J(i)=1$. 
Nevertheless, it is instructive to study  potentials \rf{Vbeta1}, \rf{Vbeta1} with arbitrary $\beta > 1$.

\begin{figure}[H]
\centering
\includegraphics[scale=0.35]{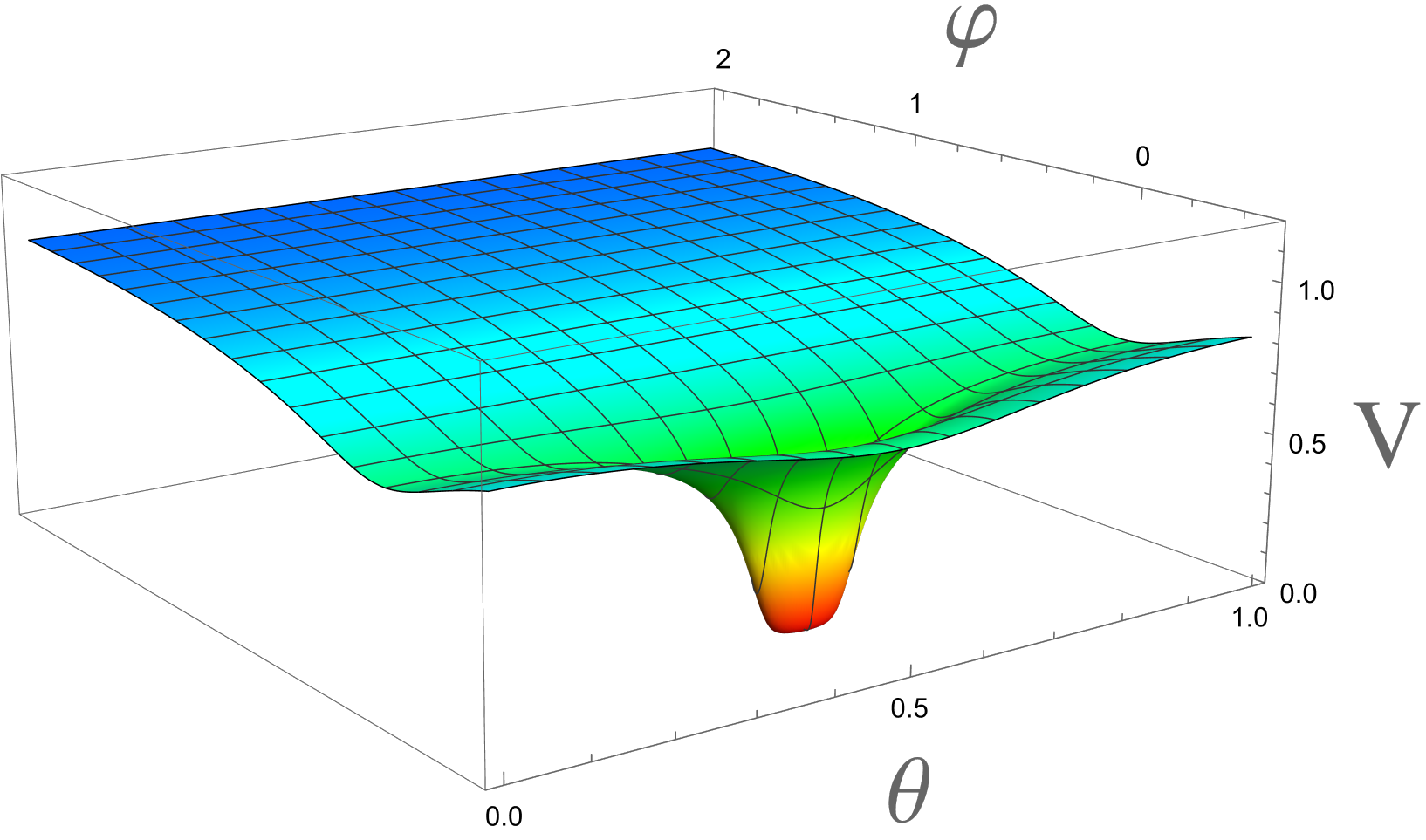}
\vskip -5pt
\caption{\footnotesize  Potential $V_2$ \rf{Vbeta2}  for $V_{0} = 1$,  $n=1$, $\alpha=1/3$, $\beta = 10$, $\tau_2= e^{\sqrt{2} \varphi}$ and $\tau_1=\theta$.  }
\label{HighPot3D}
\end{figure}

We found that the choice of $\beta \gg j(i)$ does not result in qualitative modifications of the properties of the potentials, but the choice of $\beta \ll j(i)$ may lead to a substantial change of the inflationary model. We illustrate it in Fig. \ref{HighPot3D}, which shows the potential $V_2$ \rf{Vbeta2}  for $\alpha=1/3$ and $\beta = 10$, which is 3 orders of magnitude smaller than  $j(i)$.

In this scenario, the inflationary trajectory at $\theta = 0$ and the saddle point at $\theta = \phi = 0$ becomes significantly uplifted. This leads to two potentially important effects.  First of all, if the uplifting is large enough, it may change the inflationary predictions and lead to larger values of $n_{s}$ and smaller $r$, as in the hybrid attractor scenario \cite{Kallosh:2022ggf}. Moreover, for sufficiently small $\beta > 1$, the potential of the axion field becomes very flat in the vicinity of the saddle point $\theta = \phi = 0$, and the universe may encounter a second inflationary stage when the field $\theta$ rolls from the saddle point to the minimum at $\theta =  0.5$. This may lead to nontrivial modifications of the spectrum of perturbations and even to the production of primordial black holes \cite{Kallosh:2022vha,Braglia:2022phb}.

We should say that $\beta \ll  j(i)$ is not our favorite choice. However, one should remember that we are at the very early stages of the investigation of $SL(2,\mathbb{Z})$ attractors, so we should be careful and try not to miss some interesting opportunities.

\subsection{Examples with $\eta$-function}\label{Sec:eta} 
Consider the automorphic form $\hat E_1 (\tau, \bar \tau)$ defined in eq. \rf{hatE}  and normalize it at $|\tau|=1$. $I (\tau, \bar \tau)$ here is proportional to $L_\eta$
\be 
I (\tau, \bar \tau)\equiv {\hat E_1 (\tau, \bar \tau)\over  E^0}
\, ,  \qquad \hat E_1 (\tau, \bar \tau)|_{|\tau |=1}\equiv E^0
\, , \qquad I(\tau, \bar \tau)_{|\tau|=1}=1 \ .
\label{I}\ee
At large $\tau_2$ and vanishing $\tau_1$	 the properties of $\hat E_1 (\tau, \bar \tau)$  are given in  eq. \rf{Atiyah}  so that 
\be
I (\tau, \bar \tau)|_{\tau_2 \to \infty, \tau_1=0} \to {\pi^2\over 3 E^0}\tau_2\equiv c\tau_2  \ .
\label{asympt}\ee
Our  examples will depend on  $SL(2,\mathbb{Z})$ invariant $I (\tau, \bar \tau)$ in eq. \rf{I}.

\noindent The  $SL(2,\mathbb{Z})$  invariant potentials are 
\be\label{V1}
V_1^{n}(\tau, \bar \tau)= V_0 \, \left({I (\tau, \bar \tau) - 1 \over I (\tau, \bar \tau) + 1}\right)^{n}\  ,  
\ee
\be \label{V2}
V_2^n= V_0 \left ( 1-  I^{-1} (\tau, \bar \tau)\right)^{n} \ .
\ee

These potentials vanish at the point $|\tau| =1$ when $\tau=\pm{1\over 2} + i \sqrt{3/4}$.
For  $n>1$ we see that these potentials have a minimum at $|\tau| =1$ when $\tau=\pm{1\over 2} + i \sqrt{3/4}$.

\begin{figure}[H]
\centering
\includegraphics[scale=0.37]{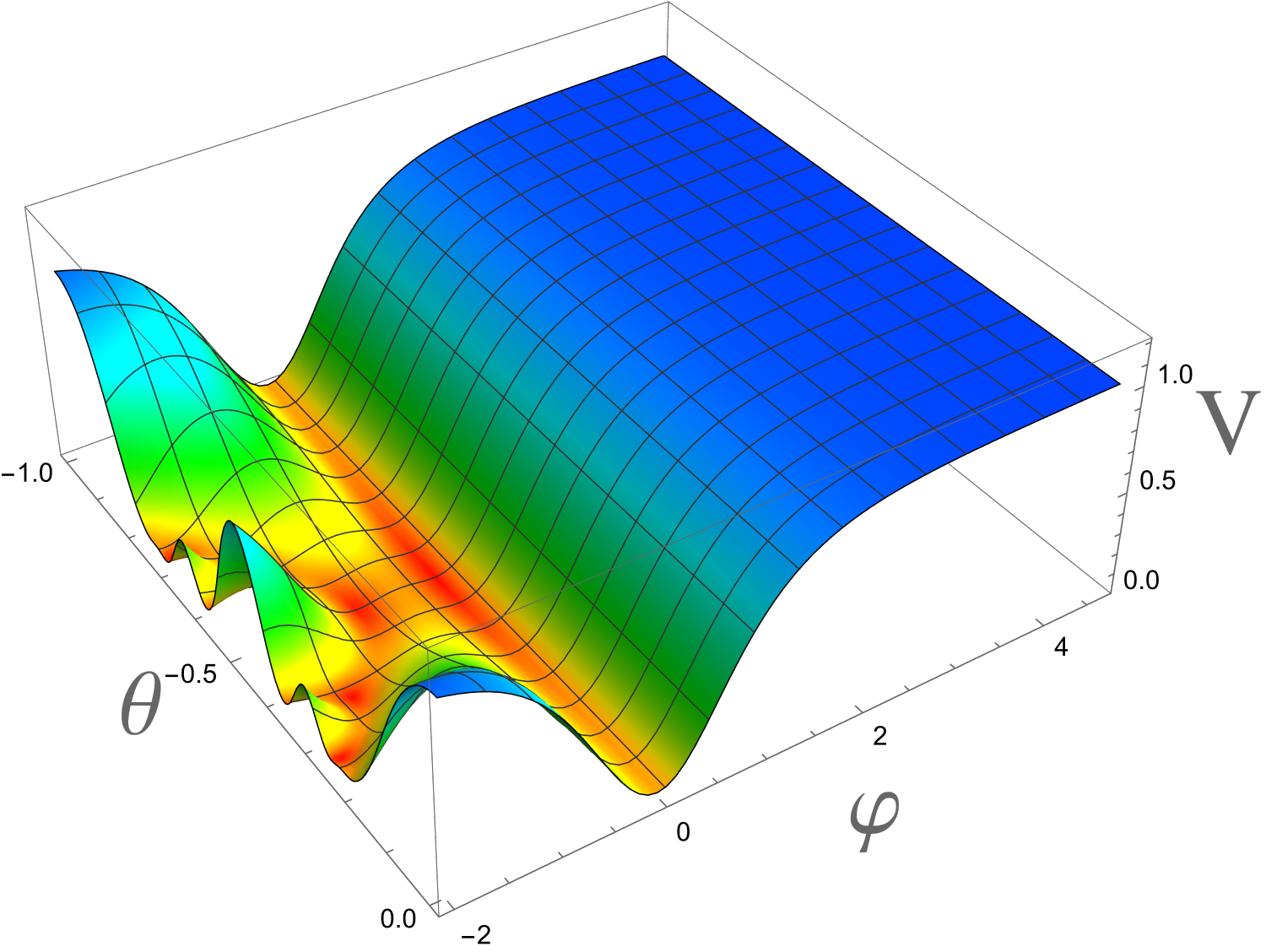}
\vskip -10pt
\caption{\footnotesize  Potential $V_2$ \rf{V2}  for $V_{0} = 1$,  $n=1$, $\alpha=1/3$, $\tau_2= e^{\sqrt{2} \varphi}$ and $\tau_1=\theta$.  }
\label{Ded2}
\end{figure}

At large $\tau_2$  with the account of eq. \rf{asympt} we find
\be
 V_1^{n}(\tau, \bar \tau)|_{\tau_2 \to \infty} \to V_0 \left (  {c\tau_2 - 1 \over  c\tau_2  + 1}\right )^{n} \ ,
\ee
and 
\be\label{D2}
V_2 (\tau, \bar \tau)|_{\tau_2 \to \infty} = V_0 \Big (1 - {1\over c\tau_2} \Big)^n  \ .
\ee
We recognize the attractor pattern studied before. The plots are very similar to the plots of the potentials with $j$-function, see e.g. Fig. \ref{Ded2} which shows the potential $V_2$ \rf{V2} for $n  = 1$.

\

\subsection{Casas-Ibanez potential \cite{Casas:2024jbw}  and its generalization}

\

\begin{figure}[H]
\centering
\includegraphics[scale=0.32]{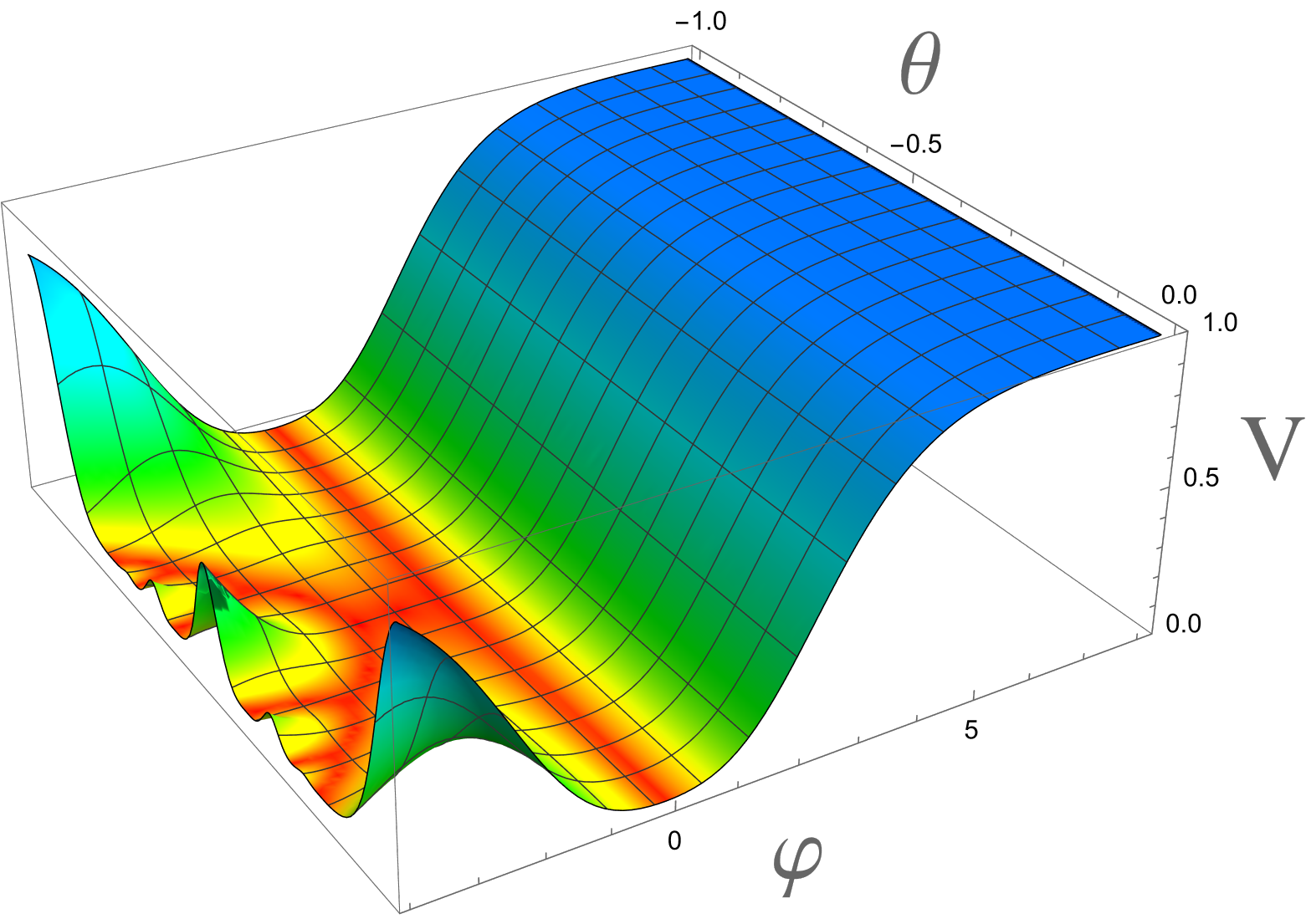}
\vskip -10pt
\caption{\footnotesize The figure presents the potential \rf{CI} for $V_0 = 1$, $\alpha=1$, $\tau_2= e^{\sqrt{2\over 3} \varphi}$, $\tilde N_0=25$  and $\tau_1=\theta$
}
\label{Ibanez}
\end{figure}

The Casas-Ibanez potential  \cite{Casas:2024jbw} depends on the ratio of two $SL(2,\mathbb{Z})$ invariants defined earlier: 
\be
V_{CI}=V_0 {L_{G_2}(\tau, \bar \tau)\over (2\pi L_{\eta} (\tau, \bar \tau) +\tilde N_0)^2} \ .
\label{CI}\ee
At $ {\rm Im}(\tau) \to \infty$, assuming that the terms ${\ln\tau_2\over \tau_2}$ in eq. \rf{as} can be ignored during inflation,
\be
V_{CI}\to V_0 \Big ({\pi\over 3} -{1\over \tau_2}\Big)^2 \ .
\ee
We have now confirmed that ignoring terms ${\ln\tau_2\over \tau_2}$ during inflation is consistent, see equations \rf{ev}, \rf{ev1}. 
 The parameter $a$ in  \cite{Casas:2024jbw} is related to $\alpha$ in eq. \rf{hyper2} as 
\be
a^2= {2\over 3\alpha} \ .
\ee
The cosmological analysis of the model \rf{CI}  was performed in  \cite{Casas:2024jbw}  by arguing that one can ignore the effect of $\tau_1$ at large $\tau_2$. Therefore the computation of inflationary slow roll parameters was performed at $\tau_1=0$.
We present the plot of such potentials in Fig. \ref{Ibanez}. Note that in this scenario, unlike in the models considered in the previous sections, the point $\theta = 0$, $\vp = 0$ is not a saddle point but a minimum with $V = 0$. 
The cosmological predictions of this model with $n=1$ were derived in \cite{Casas:2024jbw} for various $a^2= {2\over 3\alpha}$, not only for $a^2= {2\over 3}$. They coincide with general $\alpha$-attractor predictions \cite{Kallosh:2013yoa,Galante:2014ifa,Kallosh:2021mnu} given in \rf{pred}.

One can generalize the Casas-Ibanez potential \rf{CI} by taking its higher powers,
\be
V_{CI}^n=V_0 \Big ({L_{G_2}(\tau, \bar \tau)\over (2\pi L_{\eta} (\tau, \bar \tau) +\tilde N_0)^2}\Big )^n\, , \qquad n >1  \ .
\label{CIn}\ee
These potentials are also $SL(2,\mathbb{Z})$ invariant, and since all of them are $\alpha$-attractors, their inflationary predictions do not depend on $n$ for sufficiently small  $\alpha$ and large $N_{e}$. 

\section{Inflation in $SL(2,\mathbb{Z})$ invariant models \label{Sec: axions}}    
The action of the scalar fields $\vp$ and $\theta$ in hyperbolic geometry  \rf{hyper1} with $\tau= \theta+ie^{\sqrt{2\over 3\alpha} \varphi}$ is given by\footnote{Inflation and quantum fluctuations in the context of very similar models were studied in detail in \cite{Achucarro:2017ing,Linde:2018hmx}. In these models the metric of the $\theta$ field at large $\vp$ is of the form $\sim e^{2\sqrt{\frac{2}{3\alpha}}\vp}$} 
\begin{equation}\label{thetavarphi}
\int d^4x\sqrt{-g}\left(-\frac12\partial^\mu\vp\partial_\mu\vp-\frac{3\alpha}{4}e^{-2\sqrt{\frac{2}{3\alpha}}\vp}\partial_\mu\theta\partial^\mu\theta-V(\vp,\theta)\right) \ .
\end{equation}
The equations of motion for the homogeneous fields $\vp$ and $\theta$  are
\begin{equation}
\ddot\vp+3H\dot\vp-\sqrt{\frac{3\alpha}{2}}e^{-2\sqrt{\frac{2}{3\alpha}}\vp}\, \dot\theta^{2}+V_\vp=0 \ ,
\end{equation}
\begin{equation}\label{thetaeq}
\ddot\theta+3H\dot\theta-2\sqrt{\frac2{3\alpha}}\dot\vp\dot\theta+\frac{2}{3\alpha}e^{2\sqrt{\frac{2}{3\alpha}}\vp}V_\theta=0 \ .
\end{equation}

At $\vp \gg \sqrt{\alpha}$ in the slow-roll approximation one has
\begin{equation}\label{Es1}
3H\dot\vp-\sqrt{\frac{3\alpha}{2}}e^{-2\sqrt{\frac{2}{3\alpha}}\vp} \dot\theta^{2}+V_\vp=0 \ ,
\end{equation}
\begin{equation}\label{Es2}
3H\dot\theta+\frac{2}{3\alpha}e^{2\sqrt{\frac{2}{3\alpha}}\vp}V_\theta=0 \ .
\end{equation}

During inflation, the exponentially large term $e^{2\sqrt{\frac{2}{3\alpha}}\vp}V_\theta$ in \rf{Es2} could lead to a very rapid evolution of the field $\theta$, which could destabilize inflation. Fortunately, as one can see from the plots of the inflationary potentials in the previous section, these potentials are almost exactly flat with respect to $\theta$ at large positive $\vp$. A detailed investigation of this issue, to appear in a separate publication  \cite{KalLin2024}, shows that at large $\vp$ the derivatives $V_\theta$ and $V_{\theta,\theta}$ are {\it double-exponentially} suppressed by a factor $e^{-2\pi\, e^{\sqrt{\frac{2}{3\alpha}}\vp}}$. This is a very non-trivial feature of the potentials described in the previous section, which implies that the field $\theta$ practically does not change during inflation.  

In application to the models studied in the previous section, this analysis suggests the following inflationary behavior. A long stage of inflation begins at large positive values of $\vp$. For all $\vp \gg \sqrt{\alpha}$, the inflationary process practically does not depend on $\theta$ in the full range of initial values of $\theta$. Therefore, general inflationary predictions of the new class of models should coincide with general predictions of single-field $\alpha$-attractors for a large number of e-foldings: $n_{s} = 1-{2\over N_{e}}$,  $r = {12\alpha\over N^{2}_{e}}$ \cite{KalLin2024}.  A similar conclusion was reached in  \cite{Casas:2024jbw} for the model \rf{CI}  studied there.

Of course, the situation may change if one significantly changes the model parameters. For example, we found that for some (fine-tuned) values of the parameters, the height of the saddle point of the potential  \rf{Vbeta2} at $\vp = \theta = 0$ becomes very high, which may lead to a secondary stage of inflation when the field $\theta$ rolls down to the minimum at $V = 0$ at $\theta = 0.5$, see a discussion near Fig. \ref{HighPot3D}.   This regime resembles what happens in the two-field hybrid $\alpha$-attractor models, where one may encounter a combination of two subsequent inflationary regimes. This may alter their predictions and even lead to a copious production of primordial black holes  \cite{Kallosh:2022vha,Braglia:2022phb}. 

\section{\boldmath{Exploring global structure of $SL(2,\mathbb{Z})$ potentials} \label{Sec: twins}}    

For completeness, we should say a few words on the properties of the potential at large negative values of $\vp$, to make a more detailed comparison of the $SL(2,\mathbb{Z})$ attractors and the previously known classes of $\alpha$-attractors. 

As we discussed in Section 2, there are two basic classes of $\alpha$-attractors: E-models and T-models. Potentials of  E-models have a plateau at large positive values of $\vp$, and they blow up at large negative $\vp$, see Fig. \ref{EEF}. At $\alpha = 1$, these potentials coincide with the potential of the Starobinsky model. On the other hand, the potentials of  T-models depend on the absolute value of the field $\vp$ and have two symmetric plateaus at large values of $|\vp|$, see Fig. \ref{TTF}.

All $SL(2,\mathbb{Z})$ potentials  are symmetric with respect to the change $\vp \to -\vp$  at $\tau_{1} \equiv  \theta = 0$. This is a direct consequence of the inversion symmetry at $\tau_{1} = 0$. This means that  $SL(2,\mathbb{Z})$ potentials do not describe E-models and the Starobinsky model. But at first glance, these potentials do not belong to the class of T-models either.  
\begin{figure}[H]
\centering
\includegraphics[scale=0.26 
]{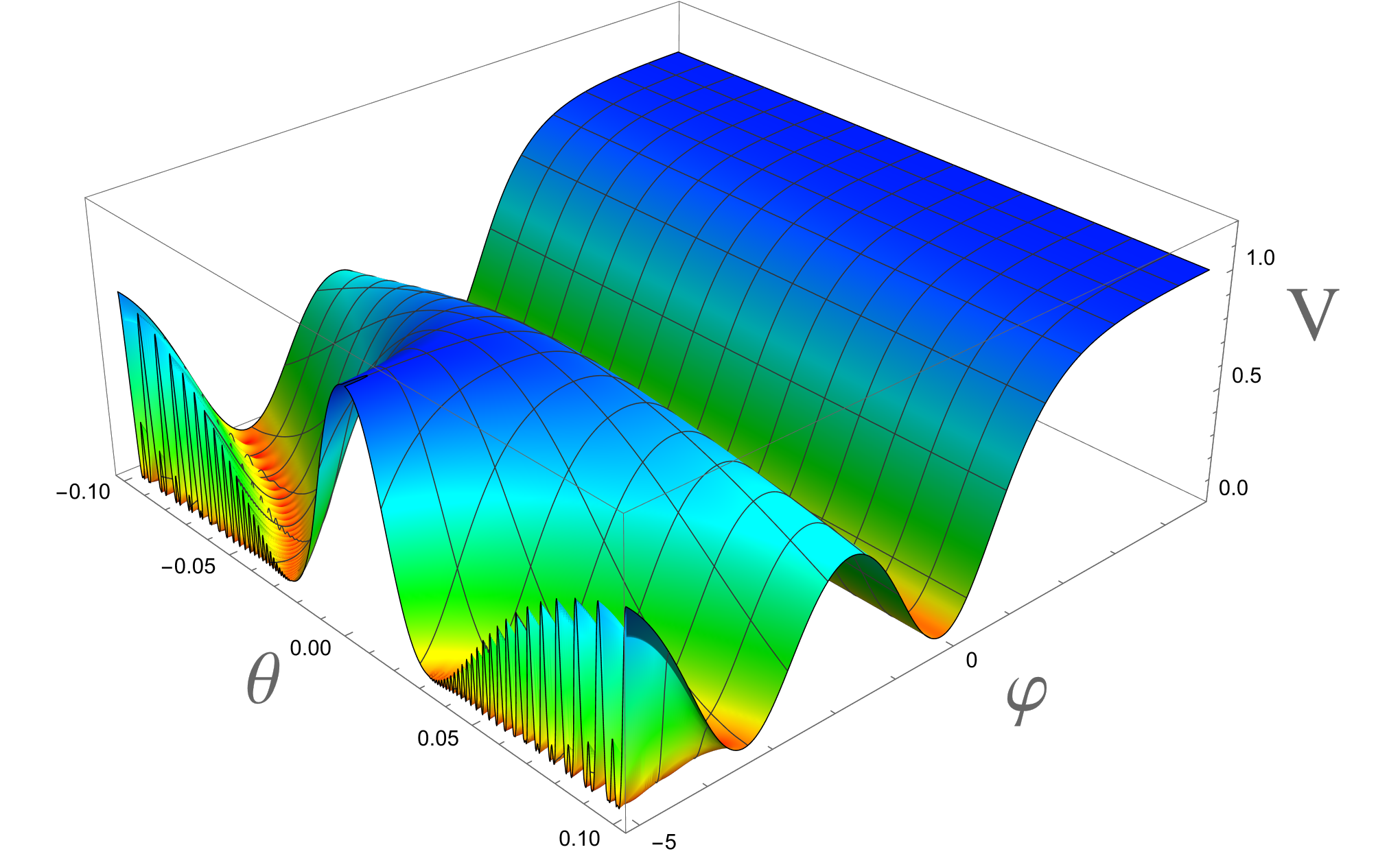}
\vskip -10pt
\caption{\footnotesize Potential $V_2$ \rf{V2}  for $V_{0} = 1$,  $n=1$, $\alpha=1/3$, $\tau_2= e^{\sqrt{2} \varphi}$ and $\tau_1=\theta$.   This figure illustrates the behavior of the potential at large negative $\vp$  for a narrow range of $\theta$ near $\theta = 0$. At $\theta = 0$, the potential is symmetric with respect to the change $\vp \to -\vp$, as in the simplest T-models. Note that the plateau at $\vp < 0$ in these coordinates ($\tau= \theta+ie^{\sqrt{2\over 3\alpha} \varphi}$) looks not like a plateau but like a ridge. As we will show in  Fig. \ref{RenPot}, 
}
\label{Negative}
\end{figure}
Indeed, in Fig. \ref{Negative}, we show the same potential $V_2$ \rf{V2} as in Fig. \ref{Ded2}, but now we concentrate on its part at large negative $\vp$, in a narrow interval of $\theta$ close to $\theta = 0$. As one can see, at large negative $\vp$ and $\theta = 0$, this potential does have an inflationary plateau. However,  in the coordinates $\tau= \theta+ie^{\sqrt{2\over 3\alpha} \varphi}$ which we used in this section, as well as in Section \ref{Sec:Inv}, this plateau looks like an infinitely long ridge. Therefore one could wonder whether an inflationary regime at $\vp < 0$ is stable.  

As a first step towards answering this question, we note that an impact of the gradients of the potential $V_{\theta}$ pushing the field $\theta$ away from the top of the ridge at  $\vp <0$ is strongly suppressed by the factor $\exp ({2\sqrt{\frac{2}{3\alpha}}\vp})$, which is {\it exponentially small} for $\vp <0$, $|\vp| \gg \alpha$, see equations \rf{thetaeq}, \rf{Es2}. This effect often leads to the {\it rolling on the ridge} regime, which was found and extensively studied in \cite{Achucarro:2017ing,Linde:2018hmx}.  This suggests that inflation along the ridge at $\vp < 0$, $\theta = 0$ is indeed possible, and $SL(2,\mathbb{Z})$ cosmological attractors discussed in this paper do belong to the class of T-models. 

\begin{figure}[H]
\centering
\includegraphics[scale=0.27 
]{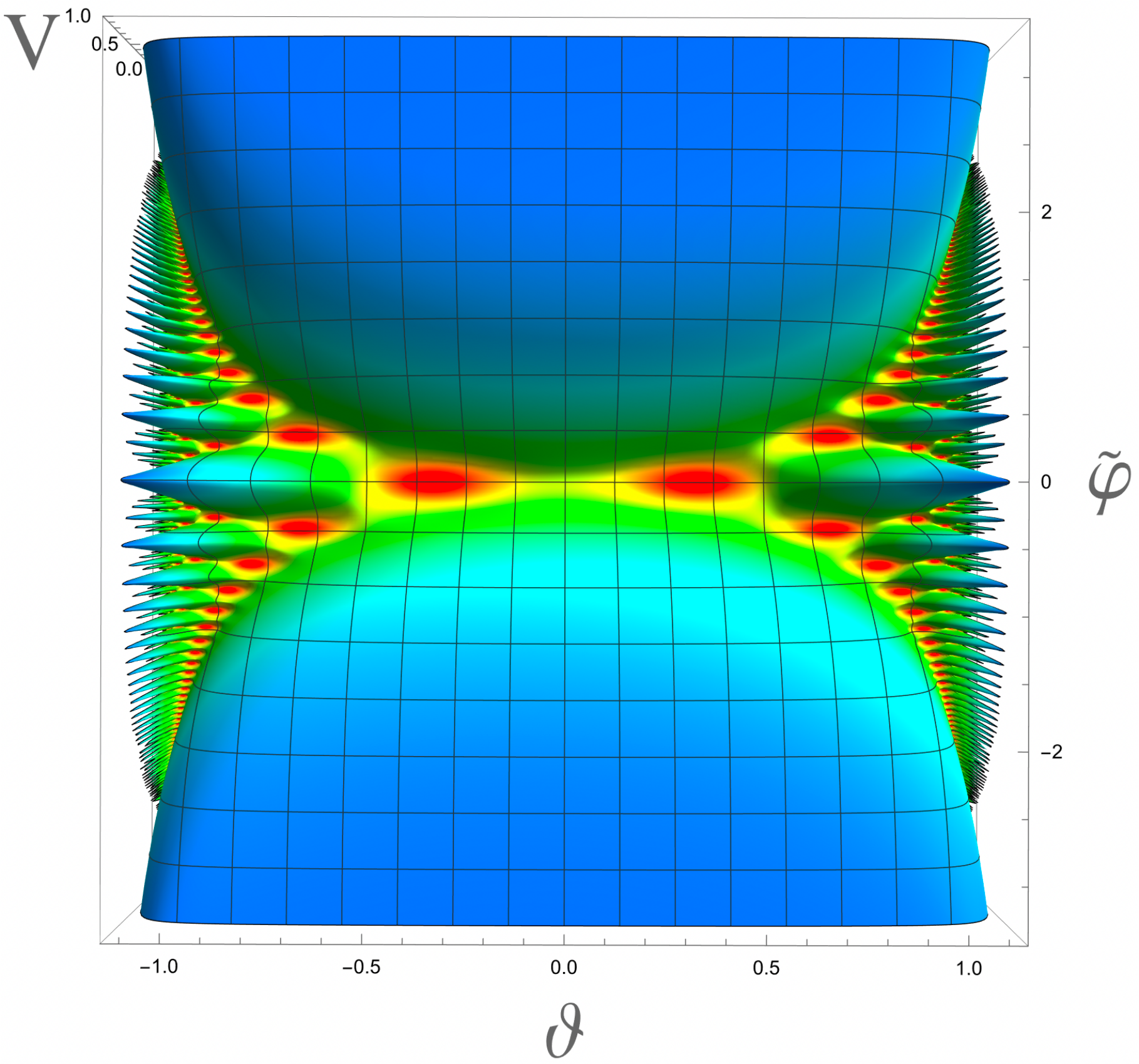}
\vskip -15pt
\caption{\footnotesize Potential $V_2$ \rf{V2}  for $V_{0} = 1$,  $n=1$, $\alpha=1/3$,  in the coordinates \rf{polar}. 
}
\label{RenPot}
\end{figure}

A more detailed description of this issue will be given in the subsequent publication  \cite{KalLin2024a}. As a preview, we present in Fig. \ref{RenPot} the same potential \rf{V2} as in Fig. 
\ref{Negative}, but in a different coordinate system, which was previously used  in \cite{Carrasco:2015rva,Carrasco:2015pla} for the description of the T-model $\alpha$-attractors:
\be\label{polar}
\tau = e^{\sqrt{2\over 3\alpha} (\tilde\vp+i \vartheta) } =  e^{\sqrt{2\over 3\alpha} \tilde\vp} \left(\sin\bigl(\sqrt{2\over 3\alpha} \vartheta\bigr) + i \cos\bigl(\sqrt{2\over 3\alpha} \vartheta\bigr) \right)\, , \quad -\pi/2 <  \sqrt{2\over 3\alpha} \vartheta < \pi/2\ .
\ee
 The kinetic term in these coordinates is
\be
{3\alpha\over 4} \, {\partial \tau \partial \bar \tau\over ({\rm Im}  \tau )^2}= {1\over 2}{ (\partial \tilde\vp)^2+ (\partial\vt)^2\over \cos^2 (\sqrt{2\over 3\alpha} \vt)} \ .
\label{kin}\ee
The inflationary plateau at $\tilde\vp > 0$ in Fig. \ref{RenPot} corresponds to the inflationary plateau at $\vp > 0$ in Fig. \ref{Negative}.  The inflationary plateau at $\tilde\vp < 0$  in Fig. \ref{RenPot} corresponds to what looked like a ridge at $\vp < 0$  in Fig. \ref{Negative}. 
In Fig. \ref{RenPot}, we show the potential ``from above'' to emphasize that these two inflationary plateaus are exact copies of each other: The potential $V_2$ \rf{V2} is manifestly symmetric with respect to the reflection symmetry $\tilde\vp \to -\tilde\vp$, as well as with respect to the reflection symmetry $\vt \to - \vt$. Note that the metric \rf{kin} has the same property: it is invariant with respect to the change $\tilde\vp \to -\tilde\vp$ and $\vt \to - \vt$.  In other words,  {\it the two inflationary plateaus shown in Fig. \ref{RenPot} are physically equivalent}. 

This means that inflation may begin not only at the plateau with $\tilde\vp > 0$, but also at the plateau with  $\tilde\vp < 0$ shown in Fig. \ref{RenPot}, and inflationary predictions \rf{pred} should not depend on the choice of the plateau, just as in T-models shown in Fig. \ref{TTF}.

This conclusion is highly nontrivial and counter-intuitive. Therefore we plan to give a more detailed investigation of the  $SL(2,\mathbb{Z})$ cosmological attractors in these coordinates in the subsequent publication  \cite{KalLin2024a}.

\section{Supersymmetric $SL(2,\mathbb{Z})$ models}\label{Sec:super}

In the framework of $\overline {D3}$ induced geometric inflation   \cite {Kallosh:2017wnt,McDonough:2016der} supergravity is defined by a function $\mathcal G = K+\ln |W|^2$ which includes in addition to our single superfield $T= -i \tau = e^{{\sqrt {2\over 3\alpha}}}-i \theta$ 
 also a superfield $X$, which is nilpotent, i.e. $X^2=0$. It is a supergravity version of the uplifting $\overline {D3}$ brane, which supports de Sitter vacuum in supergravity\cite{Bergshoeff:2015tra,Hasegawa:2015bza}.

For a half-plane variable $T$, we consider the following  K\"ahler invariant function $\mathcal G$ 
\begin{align}
\mathcal G=&-3\alpha\ln(T+\bar T)+G_{X\bar X}(T, \bar T) X\bar X +\ln |W_0+F_X \, X|^2 \ .
\end{align}
Here $X$ is a nilpotent superfield, $W_0$ is a constant defining the mass of gravitino, and $F_X$ is a constant, defining the auxiliary field vev. Following specific examples valid in $3\alpha=1$ case in \cite{Achucarro:2017ing} and $\alpha<1$ case in \cite{Yamada:2018nsk, Linde:2018hmx},  in  \cite{Kallosh:2022vha} we proposed for the case $\alpha\leq 1$  the choice
\begin{align}
G_{X\bar X}(T, \bar T)=\frac{|F_X|^2}{(T+\bar T)^{3\alpha}[ \Lambda  +V(T,\bar T)]+3|W_0|^2(1-\alpha)}\, , \qquad \Lambda =F_X^2 - 3W_0^2 \ .
\end{align}
In this case, the bosonic action following from this supersymmetric construction is 
\be
{ {\cal L} (T, \bar T)\over \sqrt{-g}} =  {R\over 2} - {3\alpha\over 4} \, {\partial T \partial \bar T\over ({\rm Re} \,  T )^2}- [\Lambda +V(T, \bar T)]  \ .
\label{hyper2}\ee
If $\Lambda =|F_X|^2 -3|W_0|^2>0$  and at the end of inflation $V(T, \bar T)=0$ there is an exit into a de Sitter vacuum. The condition 
 $\alpha \le 1$  keeps the metric  $G_{X\bar X}$ positive.

Now we take the bosonic model in eq. \rf{hyper3}  and see that we have provided a supersymmetric version of it with the choice of the function $G_{X\bar X}$ defining the relevant $V(T,\bar T)$. In this way, we have also embedded the bosonic theory into a version of supergravity with de Sitter exit of inflation.
   This gives us a supersymmetric generalization of the bosonic theories we constructed and studied in this paper and applied to cosmology.

\section{Summary}
Supergravity theories have various duality symmetries with continuous parameters, for example, $SL(2, \mathbb{R})$ symmetry. It is believed that non-perturbative quantum effects break these continuous symmetries to their discrete subgroups, 
 $SL(2, \mathbb{Z})$  in our example.  But $SL(2, \mathbb{Z})$  is expected to be preserved in exact non-perturbative theory. Recently, significant progress in constructing modular inflation was achieved in \cite{Casas:2024jbw} when the idea of $SL(2, \mathbb{Z})$  symmetry was for the first time implemented in a theory with a plateau potential.

 In this paper, we presented a broad class of $SL(2, \mathbb{Z})$ cosmological models with plateau potentials. We demonstrated that our models, as well as the model proposed in \cite{Casas:2024jbw}, belong to the general class of $\alpha$-attractors  \cite{Kallosh:2013yoa,Galante:2014ifa,Kallosh:2015zsa}, they are similar to T-models, and they have the same CMB-related observational predictions \rf{pred}. These models, as well as the string theory inspired supergravity $\alpha$-attractor models with $3\alpha=1,2,3,4,5,6,7$  \cite{Ferrara:2016fwe,Kallosh:2021vcf,Kallosh:2017ced,Kallosh:2017wnt,Kallosh:2017wku,Gunaydin:2020ric},  present a set of discrete targets for the planned CMB exploration missions such as  LiteBIRD \cite{LiteBIRD:2022cnt}. 
 
 The $SL(2, \mathbb{Z})$ cosmological models developed here show that the idea of the target space modular invariance proposed in  \cite{Ferrara:1989bc} is productive, and, moreover, falsifiable by the data. The models in  \cite{Ferrara:1989bc} have $3\alpha=n$  and the \K curvature of the ${SL(2, \mathbb{R})\over U(1)}$ coset space $\mathbb{R}_K=-{2\over n}$, where $n$ is an integer. 
 
At the time when target space modular invariance was proposed in  \cite{Ferrara:1989bc}, only supergravities with $W(\tau)$ and $K(\tau, \bar \tau)$ were known, so it was hard to find the relevant potentials. Now we can construct plateau potentials in advanced supergravity models with nilpotent superfield $X$, representing the uplifting effect of anti-D3 brane, and using the function $G_{XX}(\tau, \bar \tau)$ introduced in \cite {Kallosh:2017wnt,McDonough:2016der}. In this way, it is possible to find a consistent embedding of the new class of models in supergravity. In section \ref{Sec:super}, we show that this can be done without breaking the $SL(2, \mathbb{Z})$ symmetry, using methods developed in \cite{Achucarro:2017ing,Yamada:2018nsk,Linde:2018hmx,Kallosh:2022vha}.  

A more detailed description of the structure of the $SL(2, \mathbb{Z})$ potentials and of various features of inflation in this class of models will be contained in our subsequent publications \cite{KalLin2024,KalLin2024a}.

\section*{Acknowledgement}
We are grateful to A. Achucarro,   D. Roest, T. Wrase, and Y. Yamada for earlier work on $\alpha$-attractors, where the tools were developed, which we are also using in this project.  We are especially grateful to  G. Casas and L. Ib\'a\~nez for the useful discussion of their work \cite{Casas:2024jbw} and to R. Schimmrigk for informing us of his important work  \cite{Schimmrigk:2014ica,Schimmrigk:2016bde,Schimmrigk:2021tlv}. 
This work is supported by SITP and by the US National Science Foundation Grant   PHY-2310429.

\

\bibliographystyle{JHEP}
\bibliography{lindekalloshrefs}
\end{document}